\documentclass[11pt]{article}

\usepackage[a4paper,margin=1in]{geometry}
\usepackage{graphicx}
\usepackage{amsmath,amssymb}
\usepackage{booktabs}
\usepackage{physics}
\usepackage{placeins}
\usepackage{subcaption}
\usepackage{ragged2e}
\usepackage[numbers,sort&compress]{natbib}
\usepackage{authblk}
\usepackage{orcidlink}

\newcommand{\beq}{\begin{equation}}
	\newcommand{\eeq}{\end{equation}}
	\newcommand{\eml}{\end{mathletters}}

\newcommand{\be}{\begin{equation}}
\newcommand{\bea}{\begin{eqnarray}}
	\newcommand{\eea}{\end{eqnarray}}
\newcommand{\nn}{\nonumber\\}
\newcommand{\oh}{\frac{1}{2}}

\newcommand{\pr}{\prime}

\newcommand{\la}{\langle}
\newcommand{\ra}{\rangle}

\newcommand{\what}{\widehat}

\newcommand{\rmel}[3]{\langle #1 \| #2 \| #3 \rangle}

\begin{document}
	
	\justifying
	
	\title{Quartet Structure Above \(^{100}\)Sn and \(^{132}\)Sn Doubly Magic Isotopes}
	
	\author[1,2]{S.A. Pencu\,\orcidlink{0009-0000-6905-3782}}
	\author[1,2,3]{D.S. Delion\,\orcidlink{0000-0002-9982-0695}}
	
	\affil[1]{Horia Hulubei National Institute for R\&D in Physics and Nuclear Engineering,
		30 Reactorului, P.O. Box MG-6, RO-077125, Bucharest--M\u agurele, Rom\^ania}
	
	\affil[2]{Department of Physics, University of Bucharest,
		405 Atomi\c stilor, POB MG-11, RO-077125, Bucharest--M\u agurele, Rom\^ania}
	
	\affil[3]{Academy of Romanian Scientists,
		3 Ilfov Str., RO-050094, Bucharest, Rom\^ania}
	
	\date{}
	
	\maketitle
	

\begin{abstract}
	We calculate energy levels and B(E2) values for the \(\alpha\)-like nuclei \(^{104}\)Te and \(^{136}\)Te. Their energy structure is described within a Multi Step Shell Model (MSM) type approach by coupling proton-proton (pp), neutron-neutron (nn) and proton-neutron(pn) phonon states over the doubly magic nuclei \(^{100}\)Sn and \(^{132}\)Sn, respectively. We also compute the electric transitions for A = 102 and A = 134 Sn, Sb and Te nuclei, described within the Tamm-Dankoff Approach (TDA) with multipole-multipole residual interaction. The encountered similarities concerning the B(E2) values and wavefunctions of the coupled states corresponding to  \(^{104}\)Te and \(^{136}\)Te are analyzed.
\end{abstract}

\setcounter{equation}{0}
\renewcommand{\theequation}{1.\arabic{equation}}

%=========================================
%=========================================

\section{Introduction}

%=========================================
%=========================================

	 A consistent amount of effort and resources were dedicated in the last 
	 couple of decades to the study of proton rich nuclei and their structural 
	 characteristics \cite{Woods97,Blank08}. This region allows for a 
	 plethora of linked areas of study, ranging from $\beta$-delayed proton 
	 emissions \cite{Lund16,Del26,Bor13} to the study of rapid proton capture and two proton 
	 radioactivity, which makes it of interest to both nuclear astrophysics and 
	 nuclear physics. While important progress was made experimentally through
	 the implementation of measurement techniques such as the isotope separator
	 on-line (ISOL) and the in-flight method, there are still difficulties with probing
	 proton rich nuclei due to their extremely short half lives. From a theoretical standpoint, 
	 there have been several approaches aimed at giving a description of the nuclear structure for nuclei
	 close to the proton stability line. Results with good agreement with 
	 experimental data were achieved using mean-field and shell-model theories
	 \cite{Brown01}. 
	 
	 Measurements of isotopes above  $^{100}$Sn were performed extensively, as evidenced by \cite{San95} and the references therein. They are usually characterised by a quasiconstant energy for the $2^+$ states and increased values in the registered $\alpha$-decay reduced width.

	 The isotopic chain of Sn is one of the longest in the nuclear table and is the only one including two doubly magic nuclei $^{100}$Sn and $^{132}$Sn. 
	 The first one falls on to the proton drip line, forming also a
	local maxima of stability due to its closed shell configuration. Its spectra was
	illustrated in previous works \cite{Fae13}, using shell-model theory.
	This characteristic, together with the particle-particle 
	Tamm Dancoff Approach (pp-TDA) allows us to analyze and compare 
	potential structural differences in the excitation spectra and electric 
	transition contributions of different phonons to the total transition 
	probability. Moreover, one can evaluate the collective behaviour of the 
	resulting states illustrated by the associated TDA amplitudes. 
	
	Quartet configurations are generally described by building them from pair collective excitations \cite{Sol19,Sch21,San12,Neg12,Bar19}.
	We propose a Multi-Step-Shell-Model (MSM) approach in the 
	description of the quartet excitations, a method similarly used in other 
	papers in the description of two particle two whole configurations and  
	$\alpha$-like structures
	above $^{208}$Pb and above $^{40}$Ca in Ref. \cite{Del01,Del00}. This 
	enables us to construct the four particle phonons directly from the pair 
	states which already carry the interaction. In our description we use a
	separable multipole-multipole interaction benefiting from the clarity that 
	it provides in the analysis of multipole wave-functions. For $^{134}$Sb, 
	we employ an spin-multipole separable interaction, which compared with the modified surface delta interaction (MSDI) 
	presented in \cite{Glau67,Suh06}, allows us to account for the pairs being formed
	by particles that reside in shells with different parities.
	This asymmetry can be counted as an explanation for the negative parity states measured in the excitation spectrum of this nucleus. 
	
	Let us mention in this context the papers that employ a Shell-model code with a CD-Bonn nucleon-nucleon potential \cite{Cor06}, obtaining results in good agreement with experimental measurements.
	
	The nuclei with two protons and two neutrons above doubly magic structures, like 
	$^{40}$Ca, $^{136}$Sn and $^{208}$Pb have pronounced $\alpha$-like properties
	evidenced by spectroscopic and $\alpha$-decay data \cite{Del10}. 
	The $\alpha$-like quartet structure in medium-heavy nuclei is hindered by the Pauli principle 
	and various approaches have been proposed to account for it \cite{Flo63,Bri66,Ari71,Wil77,Gam83,Del02}.
	The $\alpha$-like structures were experimentally evidenced in light nuclei \cite{Ike68}. 
	They were extensively analyzed in Refs. \cite{Fre97,Fre07,Hor04,Fun09,Toh17}.
	
	It has been shown that the $\alpha$-like correlations in medium/heavy nuclei can also be evidenced by $\alpha$-decays.
	Although one of the first papers in theoretical nuclear physics described the 
	$\alpha$-decay in terms of the Coulomb-barrier penetration of a preformed 
	$\alpha$-particle \cite{Gam28}, the description of the $\alpha$-cluster formation by using
	nuclear structure details is still an open problem. It turns out that the quartet
	built by using two-proton and two-neutron structures given by the standard mean field
	is not enough to describe the absolute value of the decay with \cite{Man64,Ton79,Fli85,Jan83,Dod88}
	and an additional $\alpha$-structure is necessary, as it was shown in the $\alpha$-decay of $^{212}$Po \cite{Var92,Del12}.

	This paper aims to provide a simple theoretical description of the nuclear structure for the 
	$\alpha$-like nuclei $^{104}$Te and $^{134}$Te and to estimate their 
	corresponding B(E2) electrical transition probability values. These were 
	selected due to the proximity to the proton stability line of $^{104}$Te, 
	making its analysis of interest both to nuclear physics and nuclear 
	astrophysics.
	The paper is organized as follows: in Section 2 we shortly describe the used theoretical tools,
	in Section 3 we compute the energy levels and B(E2) values for A=102 and A=134 isobars,
	considered as building blocks for $\alpha$-like nuclei $^{104}$Te and $^{136}$Te. 
	In the last Section we draw conclusions.

%=========================================
%=========================================

\section{Theoretical Framework}

%=========================================
%=========================================

\setcounter{equation}{0}
\renewcommand{\theequation}{2.\arabic{equation}}
	
	\subsection{Two Particle Phonons}
	
	We will remind in this Section the main technical details used by MSM \cite{Del00}.

	In the first step we build the coupled two-particle state basis needed to compute energy values and electric transitions using the following separable Hamiltonian, as described in \cite{Del26}
	\bea
	\label{hamilt}
	H&=&\sum_{\tau j}\epsilon_{\tau j}N_{\tau j}+
	\sum_{\tau_1\tau_2}\sum_{\lambda}
	%\frac{1}{2(1+\delta_{\tau_1\tau_2})}
	G_{\lambda}(\tau_1\tau_2)
	\sum_{\mu}P^{\dag}_{\lambda\mu}(\tau_1\tau_2)P_{\lambda-\mu}(\tau_1\tau_2)(-)^{\lambda-\mu}\ ,
	\eea
	in terms of the $\lambda$-pole operators
	\bea
	\label{P}
	P^{\dag}_{\lambda\mu}(\tau_1\tau_2)=\sum_{j_1j_2}q_{\lambda}(\tau_1j_1\tau_2j_2)
	A^{\dag}_{\lambda\mu}(\tau_1 j_1\tau_2 j_2)\ ,
	\label{eq:pair_creation_op}
	\eea
defined by a coherent summation of normalized pairs
	\bea
	\label{A}
	A^{\dag}_{\lambda\mu}(\tau_1j_1\tau_2j_2)={1\over\Delta_{\tau_1j_1\tau_2j_2}}
	\left[a^{\dag}_{\tau_1j_1}\otimes a^{\dag}_{\tau_2j_2}\right]_{\lambda\mu},~~~
	\Delta_{\tau_1j_1\tau_2j_2}\equiv\sqrt{1+\delta_{\tau_1\tau_2}\delta_{j_1j_2}}\ ,
	\label{pair}
	\eea
	where $[...\otimes...]$  denotes the angular-momentum coupling.
	Here we introduced the short-hand notation for sp levels $j\equiv (\epsilon,l,j)$: sp energy, angular momentum, total spin.
	The summation in (\ref{P}) is considered over $j_1\leq j_2$ for $\tau_1=\tau_2$ and for
	all $j$ in the p-n case.
	The $q_{\lambda}$-symbol is proportional to the reduced matrix element of the standard multipole operator
	\bea
	q_{\lambda}(\tau_1j_1\tau_2j_2)&=&
	\frac{1}{\Delta_{\tau_1j_1\tau_2j_2}\widehat{\lambda}R_0^{\lambda}}
\la\tau_1 j_1 ||Q_{\lambda}||\tau_2 j_1\ra \ ,
\nn&\equiv&
	\frac{1}{\Delta_{\tau_1j_1\tau_2j_2}\widehat{\lambda}}
	\la \tau_1 j_1|\left(i\frac{R}{R_0}\right)^{\lambda}|\tau_2j_2\ra
	\la \tau_1 j_1||Y_{\lambda}(\what{R})||\tau_2j_2\ra
	\eea
	where $\widehat{\lambda}=\sqrt{2\lambda+1}$ and $R_0=1.2A^{1/3}$ is the geometrical nuclear radius.
	
	The Hamiltonian is diagonalised using the following two particle excitations of multipolarity $J$
	\bea
	\label{phonon2}
	|\tau_1\tau_2;JM n_p\ra=\Gamma^{\dag}_{JM}(\tau_1\tau_2;n_p)|0\ra=
	\sum_{j_1j_2}X_{J}(\tau_1j_1\tau_2j_2;n_p)
	A^{\dag}_{JM}(\tau_1j_1\tau_2j_2)\ ,
	\eea
	where $\tau_1\tau_2$ define proton-proton ($pp$), neutron-neutron ($nn$) and proton-neutron ($pn$) states. This two-particle creation operator defines the $n_p$-th eigenstate of the pp-TDA.

	First we will investigate pair states above $^{100}$Sn i.e. $^{102}$Te, $^{102}$Sn and $^{102}$Sb, then above $^{132}$Sn i.e. $^{134}$Te, $^{134}$Sn and $^{134}$Sb.
	
	The equation of motion 
	\bea
	\left[H,\Gamma^{\dag}_{JM}(\tau_1\tau_2;n_p)\right]=
	E_{J}(\tau_1\tau_2;n_p)
	\Gamma^{\dag}_{JM}(\tau_1\tau_2;n_p)\ ,
	\label{eq2}
	\eea
	leads to the pp-TDA system of equations
	\bea
	\label{twobody}
	\left[E_{J}(\tau_1\tau_2;n_p)-\epsilon_{\tau_1j_1}-\epsilon_{\tau_2j_2}\right]
	X_{J}(\tau_1j_1\tau_2j_1;n_p)
	=\\=g_{J}(\tau_1\tau_2)q_{J}(\tau_1j_1\tau_2j_2)\sum_{k_1k_2}
	q_{J}(\tau_1k_1\tau_2k_2)X_{J}(\tau_1k_1\tau_2k_2;n_p)\ .
	\eea
	
	\subsection{Pair B(E2) Transitions}
	The electric $\lambda$-pole transition  probabilities are given by the 
	so-called B(E$\lambda$)-values
		\begin{equation}
		B\left(E\lambda ; J \rightarrow J'\right) \equiv  \frac{1}{2 J+1}\left|\left \langle J' \left\| Q _{\lambda}\right\| J\right \rangle \right|^2
		\label{eq:def_BE2},
	\end{equation}
	where $J\equiv(J^{\pi}n)$ and $J'\equiv(J^{\pi'}n')$ are used as shorthand notations for the quantum numbers (spin, parity, eigenvalue) associated to the initial and final state and the multipole operator is given by the usual expression
	\begin{equation}
		Q_{\lambda \mu} = \sum_{\tau j_1 j_2} e(\tau) \frac{\rmel{\tau j_1}{Q_\lambda}{\tau j_2}}{\widehat{\lambda}} \left( a^\dagger_{\tau j_1} \otimes \tilde{a}_{\tau j_2} \right)_{\lambda \mu}~.
		\label{eq:operator_pair_transitions}
	\end{equation}
	We will restrict our analysis to quadrupole transitions with $\lambda=2$.
	Moving forward, using eq. \eqref{eq:pair_creation_op}, we can identify two 
	cases when computing the reduced matrix element. 
	
	(a) For $^{102}$Sn, 
	$^{134}$Sn (neutron-neutron pairs) and $^{102}$Te, $^{134}$Te 
	(proton-proton pairs), $\tau=\tau_1 = \tau_2$ the analytical expressions
	are identical
	\bea
	\rmel{J}{Q_\lambda}{J'} &=& 4e(\tau) \sum_{j_1<j_2}\sum_{j_1'<j_2} (-)^{j_1' - j_2 + J + \lambda} \widehat{J} \widehat{J'} 
	\nn&\times&
		\frac{X(\tau j_1\tau j_2) X'(\tau j_1'\tau j_2)}{\Delta_{j_1j_2} \Delta_{j_1'j_2}} \rmel{\tau j_1}{Q_\lambda}{\tau j_1'} \mathcal{W}\left( J j_1 J' j_1'; j_2 \lambda \right)
	\nn
	&&\tau=p,~n~,
	\eea
	where $\mathcal{W}$ denotes the Racah recoupling symbol.
	
	(b) In the case where the pair is formed by a proton and a 
	neutron, corresponding to $^{102}$Sb and $^{134}$Sb, the derivation 
	is similar, as well as the analytical expression, when accounting for the contribution of each isospin.
	
\begin{equation}
	\begin{aligned}
		\rmel{J}{Q_\lambda}{J'} &= 2e(p) \sum_{j_1j_2}\sum_{j'_1}
		(-)^{j_1' - j_2 + J + \lambda}
		\widehat{J} \widehat{J'}
		X(p j_1n j_2) X'(p j_1'n j_2)
		\rmel{j_1}{Q_\lambda}{j_1'}
		\mathcal{W}\left( J j_1 J' j_1'; j_2 \lambda \right) \\
		&\quad + 2e(n) \sum_{j_1j_2}\sum_{j'_1}
		(-)^{j_1' - j_2 + J + \lambda}
		\widehat{J} \widehat{J'}
		X(n j_1p j_2) X'(n j_1'p j_2)
		\rmel{j_1}{Q_\lambda}{j_1'}
		\mathcal{W}\left( J j_1 J' j_1'; j_2 \lambda \right)
	\end{aligned}
\end{equation}

	\subsection{Four Particle Phonons}
        In the second step we evaluate
	the correlated four-particle eigenstates $n_q$ of multipolarity $I$, defined as follows
	\bea
	|In_q\ra=Q^{\dag}_{I}(n_q)|0\ra\ ,
	\eea
	where the four-particle (quartet) creation operator has two main components:
	\bea
	Q^{\dag}_{I}(n_q)&\equiv&
	\sum_{Jn_pKr_p} X(ppJn_p;nnKr_p;I n_q)
	\left[\Gamma^{\dag}_{J}(pp;n_p)\otimes\Gamma^{\dag}_{K}(nn;r_p)
	\right]_{I}
	\nn
	&+&\sum_{Jn_pKr_p} X(pnJn_p;pnKr_p;In_q)
	\left[\Gamma^{\dag}_{J}(pn;n_p)\otimes\Gamma^{\dag}_{K}(pn;r_p)
	\right]_{I}\ .
	\label{fun4}
	\eea
	
	\noindent The above two terms correspond to the couplings 
	
	(a) $(^{102}{\rm Te}\otimes{^{102}{\rm Sn}})_{In_q}$ and $(^{102}{\rm Sb}\otimes{^{102}{\rm Sb}})_{In_q}$ and
	
	(b) $(^{134}{\rm Te}\otimes{^{134}{\rm Sn}})_{In_q}$ and $(^{134}{\rm Sb}\otimes{^{134}{\rm Sb}})_{In_q}$, respectively.

	By using the short-hand notations $J\equiv(\tau_1\tau_2Jn_p)$
	and $I\equiv(In_q)$, the expression \eqref{fun4} can be written as
	
	\bea
	Q^{\dag}_{I}=\sum_{JK}X(JK)
	\left[\Gamma^{\dag}_{J}\otimes\Gamma^{\dag}_{K}\right]_{I}\ .
	\label{fourpar}
	\eea
	The quartet TDA (q-TDA) equation of motion to obtain the four-particle system is given by
	\bea
	\left[H,Q^{\dag}_{I}\right]&=&E_{I}Q^{\dag}_{I}.
	\label{eq4}
	\eea
	By using pp-TDA equation (\ref{eq2}) and the symmetrised double commutator one obtains
	the following system of q-TDA equations
	\bea
	\sum_{J'K'}H_I(JK;J^{\pr}K^{\pr})
	X(J^{\pr}K^{\pr})=E_{I}
	\sum_{J'K'}N_I(JK;J^{\pr}K^{\pr})
	X(J^{\pr}K^{\pr})\ ,
	\label{syst3}
	\eea
	where the metric matrix is defined in terms of the following overlap
	\bea
	N_I(JK;J^{\pr}K^{\pr})\equiv
	\la 0|Q_{I}(JK)
	Q^{\dag}_{I}(J^{\pr}K^{\pr})|0\ra\ .
	\eea
	It turns out that the Hamiltonian matrix is proportional to the metric matrix
	and contains only pp-TDA pair energies. The derivation of the norm and hamiltonian matrix are given in Ref \cite{Del00}
	\bea
	H_I(JK;J^{\pr}K^{\pr})&\equiv&
	\oh\la0|\left[Q_{I}(JK),H,
	Q^{\dag}_{I}(J^{\pr}K^{\pr})\right]|0\ra
	\nn
	&=&\oh(E_{J}+E_{K}+E_{J^{\pr}}+E_{K^{\pr}})
	N_I(JK;J^{\pr}K^{\pr})\ .
	\eea
	
	\subsection{Quartet B(E2) Transitions}
	
	In deriving the reduced matrix element (RME), associated with quartet 
	electric transitions, we can neglect the cross ppnn-pnpn terms because their 
	corresponding amplitudes are relatively small. This leads us to the expression
\begin{equation}
	\begin{aligned}
		\langle I'M'|Q_{\lambda}|IM\rangle
		&=
		\sum_{K J_1 J_1'} \sum_{\tau_1 \neq \tau_2}
		e(\tau_1)\,
		X'(\tau_1 \tau_1 J'n_p';\tau_2 \tau_2 K'n_p';I'n_q')
		X(\tau_1 \tau_1Jn_p;\tau_2 \tau_2 Kn_p;In_q)
		\\
		&\times
		(-)^{J_1+J_2-I'+\lambda}
		\hat I'\hat I\,
		\langle J_1'||Q_\lambda||J_1\rangle
		\mathcal{W}(I'J_1'IJ_1;K\lambda)
		\\
		&+
		\sum_{K J_1 J_1'} \sum_{\tau_1 \neq \tau_2}
		e(\tau_1)\,
		X'(\tau_1 \tau_2 J'n_p';\tau_1 \tau_2 K'n_p';I'n_q')
		X(\tau_1 \tau_2 Jn_p; \tau_1 \tau_2 Kn_p;In_q)
		\\
		&\times
		(-)^{J_1+J_2-I'+\lambda}
		\hat I'\hat I\,
		\langle J_1'||Q_\lambda||J_1\rangle
		\mathcal{W}(I'J_1'IJ_1;K\lambda).
	\end{aligned}
\end{equation}

%=========================================
%=========================================

\section{Numerical Application}

%=========================================
%=========================================

\setcounter{equation}{0}
\renewcommand{\theequation}{3.\arabic{equation}}

We first analyzed pp-TDA excitations energies and B(E2)-values and then similar q-TDA quantities. The energy eigenvalues and electric transition probabilities reported here correspond to the first eigenstates. In the following tables, the eigenvalue index is denoted by a subscript attached to each state $n_i^{\pm}$.

	\subsection{Pair Energies} 
		
		We start by creating a sp basis with eigenstates 
		of a spherical Woods-Saxon mean field with standard 
		universal parametrization \cite{Cwi87}.
		The sp quantum numbers and energies within one major shell above $Z=N=50$ are given
		in Table 1 for $^{100}$Sn.
		The same quantities above $Z=50~,N=82$ are given Table 2 for $^{132}$Sn.
		 
		\begin{table}[h]
			\centering
			\label{tab:100sp}
			\caption{Single-particle energies for $^{100}$Sn, protons (left) and neutrons (right)} 
			\begin{minipage}{0.45\textwidth}
				\begin{tabular}{ccccc}
					\toprule
					No. & l & j & Energy [MeV] \\
					\midrule
					1 & 2 &$5/2 $& 3.450 \\
					2 & 0 &$1/2 $& 5.090 \\
					3 & 4 &$7/2 $& 5.310 \\
					4 & 2 &$3/2 $& 6.160 \\
					5 & 5 &$11/2$ & 6.660 \\
					\bottomrule
				\end{tabular}
			\end{minipage}
			\hfill
			\begin{minipage}{0.45\textwidth}
				\begin{tabular}{cccc}
					\toprule
					No. & l & j & Energy [MeV] \\
					\midrule
					1 & 2 &$5/2 $& -12.690 \\
					2 & 4 &$7/2 $& -11.320 \\
					3 & 0 &$1/2 $& -10.550 \\
					4 & 2 &$3/2 $& -9.890 \\
					5 & 5 &$11/2$ & -9.820 \\
					\bottomrule
				\end{tabular}
			\end{minipage}
		\end{table}
		
		\begin{table}
			\centering
			\label{tab:132sp}
			\caption{Single-particle energies for $^{132}$Sn, protons (left) and neutrons (right)}
			\begin{minipage}{0.45\textwidth}
				\begin{tabular}{ccccc}
					\toprule
					No. & l & j & Energy [MeV] \\
					\midrule
					1 & 4 & 7/2 & -9.640 \\
					2 & 2 & 5/2 & -9.540 \\
					3 & 5 & 11/2 & -7.370 \\
					4 & 0 & 1/2 & -7.030 \\
					5 & 2 & 3/2 & -6.790 \\
					\bottomrule
				\end{tabular}
			\end{minipage}
			\hfill
			\begin{minipage}{0.45\textwidth}
				\begin{tabular}{ccccc}
					\toprule
					No.& l & j & Energy [MeV] \\
					\midrule
					1 & 3 & 7/2 & -2.580 \\
					2 & 1 & 3/2 & -1.350 \\
					3 & 5 & 9/2 & -0.880 \\
					4 & 1 & 1/2 & -0.630 \\
					5 & 3 & 5/2 & -0.200 \\
					6 & 6 & 6 1/2 & 0.200 \\
					\bottomrule
				\end{tabular}
			\end{minipage}
			
		\end{table}

		This sp basis is used to build pair states by using pp-TDA in order to 
		obtain the amplitudes and energy eigenvalues. The obtained results 
		for $^{102}$Te, $^{134}$Te and their isobars are tabulated in Table 
		\ref{tab:102_energies} and Table \ref{tab:134_energies}, respectively.
		Here we estimated the excitation energy by the difference
\bea
E_{ex}(J)=E_J-E_0~.
\eea		
Let us mention that we used the strength conefficients of the hamiltonian (\ref{hamilt}) $G_{\lambda}(\tau_1\tau_2)$
given in Table \ref{strength}.
		For A = 134 isobars, we remark the unnatural parity states of the 
		odd-odd nucleus \(^{134}\)Sb due to the coupling of proton and neutron states
		with different parities. In general, one can describe such states 
		using a modified surface delta interaction, as presented in 
		\cite{Glau67}. In this case, however, it is not a suitable description due 
		to the manner in which the pairs are created, where the protons and 
		neutrons that form them originate from shells with different parities. 
		Recent papers describe these collective states using a shell model 
		framework with CD-Bonn nucleon-nucleon potential \cite{Cor06}, with 
		good agreement with the measured data.
		
		We opted in this case for a separable interaction where we have constructed the 
		nuclear matrix elements by using the spin-orbit interaction 
		\bea
		T_{\lambda}(l_s)=\left[Q_{l_s} \otimes \sigma \right]_\lambda\equiv
		R^{l_s}\left[i^{l_s} Y_{l_s} \otimes \sigma \right]_\lambda~,
		\eea
		in order to connect sp orbitals with different parities. Furthermore we 
		noticed that the nucleon-nucleon interaction strength for 0\(^-\) state should be   
		different from the other spins in order to obtain a reasonable agreement
		with experimental energy values.
		
		\begin{table}[h]
			\centering
	\caption{Calculated pair excitation energies of the lowest lying eigenstates for A = 102 isobars compared with experimental determinations and corresponding wavefunction structure for the pair with the maximal TDA amplitude}
		\begin{tabular}{ccccccc}
			\toprule
			Nucleus & Multipole & Energy (Th) & Energy (Exp) & Pair Structure & Max. Amplitude \\
			& & [MeV] & [MeV]& \\
			\midrule
			\(^{102}\)Te & 0\(_1^+\) & 0.000 & 0.000 & d$_{5/2}$ d$_{5/2}$ & 0.952 \\
			& 2\(_1^+\) & 0.691 & 0.660 & d$_{5/2}$ d$_{5/2}$ & 0.986 \\             
			& 4\(_1^+\) & 0.680 & 1.274 & d$_{5/2}$ d$_{5/2}$ &  0.983\\             
			& 6\(_1^+\) & 1.957 & 1.057 & f$_{7/2}$ d$_{5/2}$ & 0.991 \\             
			\(^{102}\)Sn & 0\(_1^+\) & 0.000 & 0.000 & d$_{5/2}$ d$_{5/2}$ & 0.865\\
			& 2\(_1^+\) & 1.442 & 1.472 & d$_{5/2}$ d$_{5/2}$  & 0.989\\
			& 4\(_1^+\) & 1.510 & 1.969 & d$_{5/2}$ d$_{5/2}$ &  0.994\\
			& 6\(_1^+\) & 2.601 & 2.122 & f$_{7/2}$ d$_{5/2}$ &  0.995\\         
			\(^{102}\)Sb & 0\(_1^+\) & 0.000 & 0.000 & d$_{5/2}$ d$_{5/2}$ & 0.999\\       
			& 2\(_1^+\) & 0.122 & 0.150 & d$_{5/2}$ d$_{5/2}$ &  0.999 \\
			& 4\(_1^+\) & 0.129 & 0.310 & d$_{5/2}$ d$_{5/2}$ &  0.999\\          
			& 6\(_1^+\) & 1.960 & 0.531 & f$_{7/2}$ d$_{5/2}$ &  0.999\\       
			\bottomrule
		\end{tabular}
		\label{tab:102_energies}
	\end{table}

	\begin{table}[h]
		\centering
\caption{ Calculated pair excitation energies of the lowest lying eigenstates for A = 134 isobars compared with experimental determinations and corresponding wavefunction structure for the pair with the maximal TDA amplitude}
		\begin{tabular}{ccccccc}
			\toprule
			Nucleus & Multipole & Energy (Th) & Energy (Exp) & Max. Pair & Max. Amplitude \\
			& & [MeV] & [MeV]& \\
			\midrule
			\(^{134}\)Te & 0\(_1^+\) & 0.000 & 0.000 & f$_{7/2}$ f$_{7/2}$ &0.772\\
			& 2\(_1^+\) & 1.310 & 1.279 & f$_{7/2}$ f$_{7/2}$ &0.978\\
			& 4\(_1^+\) & 1.356 & 1.576 & f$_{7/2}$ f$_{7/2}$ &0.996\\
			& 6\(_1^+\) & 1.364 & 1.691 & f$_{7/2}$ f$_{7/2}$ &0.998\\
			\(^{134}\)Sn & 0\(_1^+\) & 0.000 & 0.000 & f$_{7/2}$ f$_{7/2}$ &0.838\\
			& 2\(_1^+\) & 1.506 & 0.725 & f$_{7/2}$ f$_{7/2}$ &0.940\\
			& 4\(_1^+\) & 1.472 & 1.073 & f$_{7/2}$ f$_{7/2}$ &0.894\\
			& 6\(_1^+\) & 1.332 & 1.247 & f$_{7/2}$ f$_{7/2}$ &0.851\\
			\(^{134}\)Sb & 0\(_1^-\) & 0.000 & 0.000 & d$_{5/2}$ e$_{5/2}$ &0.943\\
			& 1\(_1^-\) & 0.150 & 0.13 & d$_{5/2}$ f$_{7/2}$ &0.656\\
			& 7\(_1^-\) & 0.979 & 0.279 & d$_{5/2}$f$_{7/2}$ &0.656\\
			& 2\(_1^-\) & 0.541 & 0.331 & d$_{5/2}$ f$_{7/2}$ &0.656\\
			& 3\(_1^-\) & 0.895 & 0.384 & d$_{5/2}$ f$_{7/2}$ &0.656\\
			& 5\(_1^-\) & 0.995 & 0.441 & d$_{5/2}$ f$_{7/2}$ &0.656\\
			& 4\(_1^-\) & 0.876 & 0.555 & d$_{5/2}$ f$_{7/2}$ &0.883\\
			& 6\(_1^-\) & 0.802 & 0.617 & d$_{5/2}$ f$_{7/2}$ &0.981\\
			\bottomrule
		\end{tabular}
		\label{tab:134_energies}
	\end{table}
	
	Let us notice that pairs with the largest contribution to the final excited states are 
	built by coupling sp states with identical angular momenta and spins.
	Apart from the negative parity states, we also notice a more 
	pronounced collective character in the excitation spectra of $^{134}$Sb. 
	This is attributed the coupling of particles from major shells with different parities.
	
	\begin{table}
		\centering
		\caption{Strength coefficients that best reproduce overall energy eigenvalues \\ 
		for A=102 and A=134 isobars}
		\begin{minipage}{0.45\textwidth}
			\begin{tabular}{cccc}
				\toprule
				Nucleus & $\tau_1-\tau_2$ & $G_{\lambda}(\tau_1\tau_2)$ \\
				\midrule
				$^{102}$Te & p-p & -2.2\\
				$^{102}$Sn & n-n & -3.1\\
				$^{102}$Sb & p-n & -0.3\\
				\bottomrule
			\end{tabular}
		\end{minipage}
		\hfill
		\begin{minipage}{0.45\textwidth}
			\begin{tabular}{cccc}
				\toprule
				Nucleus & $\tau_1-\tau_2$ & $G_{\lambda}(\tau_1\tau_2)$ \\
				\midrule
				$^{134}$Te & p-p & -2.0\\
				$^{134}$Sn & n-n & -2.4\\
				$^{134}$Sb & p-n & -7.3\\
				\bottomrule
			\end{tabular}
		\end{minipage}
	\label{strength}
	\end{table}

	\FloatBarrier

\subsection{Pair Transitions}

	We have tabulated the calculated B(E2) values for the previously analyzed 
	nuclei. For A = 102 isobars, the presented values do not have 
	an experimental comparison due to extensive difficulties regarding their 
	measurements. 	
	Concerning A = 134 isobars we estimated the effective charges for protons and neutrons by computing the ratio between experimental data, where available, and theoretical estimates for nuclei with proton-proton and neutron-neutron pair configurations. We have found the nucleon effective charges $e_p = 1.31 e$ and $e_n =  0.33 e$, which are consistent with the systematic presented in \cite{Del13} giving 
	$e_p = 1.20 e$ and $e_n =  0.20 e$.
	This enables us to provide estimated values for $^{134}$Sb and $^{102}$Sb.

\begin{table}[h]
\centering
\caption{
Calculated B(E2) values for A=102 isobars
}
\label{tab:102_transitions}
\begin{tabular}{cccc}
\toprule
Nucleus & Transition Type & B(E2) (Th) & B(E2) (Th) \\
& & [e$^2$ fm$^4$] & [W.u.] \\
\midrule

\(^{102}\)Te & 2$_1^+$ — 0$_1^+$ & 149.452 & 3.169 \\
& 4$_1^+$ — 2$_1^+$ & 93.487 & 1.981 \\
& 6$_1^+$ — 4$_1^+$ & 0.360 & 0.009 \\

\(^{102}\)Sn & 2$_1^+$ — 0$^+$ & 8.830 & 0.187 \\
& 4$_1^+$ — 2$_1^+$ & 5.802 & 0.123 \\
& 6$_1^+$ — 4$_1^+$ & 0.004 & 0.000 \\

\(^{102}\)Sb & 2$_1^+$ — 0$_1^+$ & 132.447 & 2.807 \\
& 4$_1^+$ — 2$_1^+$ & 80.438 & 1.704 \\
& 6$_1^+$ — 4$_1^+$ & 0.000 & 0.000 \\

\bottomrule
\end{tabular}
\end{table}

\begin{table}[h]
\centering
\caption{Calculated and measured (where available) B(E2) values for A=134  corresponding to first eigenvalue index states}
\label{tab:134_transitions}
\begin{tabular}{ccccc}
\toprule
Nucleus & Transition Type & B(E2) (Th) & B(E2) (Th) & B(E2) (Exp)\\
& & [e$^2$ fm$^4$] & [W.u.] & [W.u.]\\
\midrule
\(^{134}\)Te & 2$_1^+$ — 0$_1^+$ & 68.730 & 1.020 & 1.02\\
& 4$_1^+$ — 2$_1^+$ & 54.520 & 0.807 & - \\
& 6$_1^+$ — 4$_1^+$ & 12.648 & 0.189 & -\\

\(^{134}\)Sn & 2$_1^+$ — 0$_1^+$ & 18.720 & 0.284 & 0.284\\
& 4$_1^+$ — 2$_1^+$ & 20.321 & 0.299 & -\\
& 6$_1^+$ — 4$_1^+$ & 13.188 & 0.194 & -\\

\(^{134}\)Sb & 2$_1^-$ — 0$_1^-$ & 0.562 & 0.009 & -\\
& 4$_1^-$ — 2$_1^-$ & 60.565 & 0.891 & - \\
& 6$_1^-$ — 4$_1^-$ & 29.440 & 0.432 & - \\
\bottomrule
\end{tabular}
\end{table}
	
	\FloatBarrier
	
	\subsection{Quartet Energies}
	
		We estimated the eigenvalues associated with the quartet structure description 
		by using the pair states as building blocks and recoupling 
		them to form the quartet basis, according to the MSM method.  The 
		energy eigenvalues and q-TDA amplitudes were then determined by solving 
		the resulting q-TDA equations \eqref{syst3}. The procedure to solve this
		system of equations is described in \cite{Del00}. First we diagonalized the norm matrix
		by keeping the eigenvalues larger than $N_{min}=0.4$ which do not violate the Pauli principle.
		The eigenvalues of the metric matrix versus the eigenvalue number are plotted 
		in Fig. 1 for $I=2^+$.
		Then we used the eigenstates of the norm matrix in order to built a hermitean hamiltonian
		matrix which can be diagonalized by using the standard subroutine.
		We have been displayed the results in Table \ref{tab:quartet_energies}. Here we can see
		that the largest component have a pure pair pp-nn structure, where the quartet spin
		is given by the proton pair for both Te isotopes. Theoretical energies for $^{136}$Te
		are in a reasonable agreement with experimental values.
		
		\begin{figure}
			\centering
			\includegraphics[width=9cm]{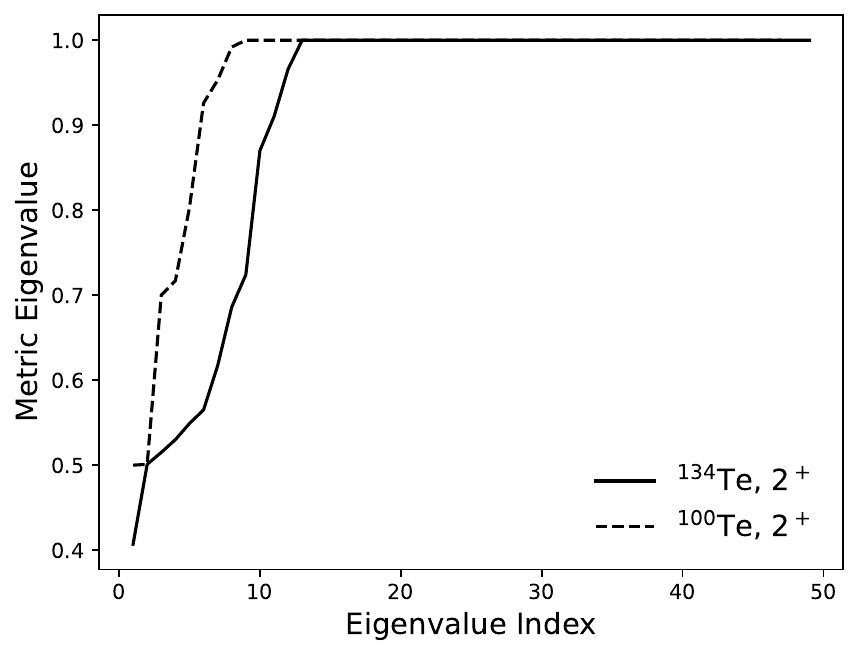}
			\label{fig:metric_eig}
			\caption{The eigenvalues of the metric matrix for $^{104}$Te (dashed line) and $^{136}$Te (solid line) versus the eigenvalue index.}
		\end{figure}

\begin{table}[h]
\centering
\caption{
Quartet pair structure and excitation energies compared to the measured data (where possible) for $^{104}$Te and $^{136}$Te first eigenstates. Also presented are the pair angular momenta and maximal quartet TDA amplitude.
}
\begin{tabular}{ccccccc}
\toprule
Nuclei & Multipole & Quartet & Max. Ampl. & $J_{pp}-J_{nn}$ & Energy (Th) & Energy (Exp)
\\
& & structure & & &[MeV]&[MeV]\\
\midrule
$^{104}$Te & 0$_1^+$ & pp-nn  & 1.00 & 0-0 & 0.000 & - \\
& 2$_1^+$ &pp-nn &1.00 & 2-0 & 0.691  & - \\
& 4$_1^+$ &pp-nn& 1.00& 4-0 & 0.680  & - \\
& 6$_1^+$ &pp-nn& 1.00& 6-0 & 1.957  &  - \\
$^{136}$Te & 0$_1^+$ & pp-nn & 1.00& 0-0&0.000 & 0.000 \\
& 2$_1^+$ &pp-nn &1.00&2-0 & 1.321  &0.606 \\
& 4$_1^+$ &pp-nn &1.00&4-0 & 1.360  &1.030 \\
& 6$_1^+$ &pp-nn &1.00&6-0 & 1.365  &1.382 \\
\bottomrule
\end{tabular}
\label{tab:quartet_energies}
\end{table}		
		
		For $^{104}$Te we have a lack of experimental data due to its 
		proximity to the proton drip line. However, we have obtained a reasonable 
		agreement between our results and measurements for the more stable 
		$^{136}$Te, thus allowing us to be carefully confident in predicting the 
		excitation spectrum for its correspondent.
		
	\subsection{Quartet B(E2) Transitions}
		Finally we analyzed B(E2) values from $\alpha$-like nuclei $^{104}$Te and $^{136}$Te.
		Our analysis evidenced the exclusive character of the pp-nn quartet types contributions
		to electric transitions in both cases, 
		as it is illustrated in Table \ref{tab:quartet_transitions}. This result is 
		interesting due to the differences in the structure of the pairs
		contributing to quartets. It turns out that in the case $^{104}$Te proton-neutron
		correlations do not play an important role in both the wave function structure and
		electromagnetic transitions, as expected.
		The results are similar to the other case of $^{136}$Te, where the protons and 
		neutrons originate from shells with different parities, influencing, as shown before, 
		multipole parity and transition probabilities for the $^{134}$Sb nucleus.
		Let us mention that a similar conclusion concerning $\alpha$-decays was obtained
		in \cite{Bar16}. Here the $\alpha$-cluster component of the potential
		was described by a common systematics in terms of the reduced width above 
		$^{100}$Sn, as well as $^{208}$Pb.
		
		\begin{table}[h]
			\centering
			\caption{
				Calculated B(E2) transition values for $^{104}$Te and $^{136}$Te
			}
			\begin{tabular}{cccccccc}
				\toprule
				Nucleus & Transition & Quartet &Max. Ampl. & B(E2) [W.u.]\\
				\midrule
				$^{104}$Te & 2$_1^+$ — 0$_1^+$ & pp-nn &1.00 & 0.20\\
				& 			      & pp-nn &1.00  & \\
				$^{136}$Te & 2$_1^+$ — 0$_1^+$ & pp-nn &1.00 & 0.11\\
				& 			        & pp-nn &1.00 & \\
				\bottomrule
			\end{tabular}
			\label{tab:quartet_transitions}
		\end{table}
		
		\FloatBarrier
		
		While we do not observe any collective effects for low eigenvalue states, the collectivity increases for higher eigenvalues. This is evidenced by the reduction of the maximal amplitude for states with eigenvalue indexes over 30 and correlated by larger off-diagonal terms in the hamiltonian matrix. Similar effects were noticed for $^{212}$Po \cite{Del00}, where a larger initial basis was used.

\section*{Conclusions}

	In this paper we have analyzed the low lying states and electrical 
	quadrupole transition probabilities for the $\alpha$-like nuclei $^{104}$Te and 
	$^{134}$Te. The single particle states, used later as a basis for the pair 
	states, were determined using a Woods-Saxon mean field with universal 
	parametrisation.  We employed a pp-TDA formalism together with a 
	separable hamiltonian in order to describe the multipole excitation 
	spectra in A=102 and A=134 isobars. 
	The MSM approach allowed us to describe the quartet states by coupling the pp-TDA pair states. 
	We have found that, in spite of the differences between the 
	pair structure of the two nuclei $^{104}$Te and $^{134}$Te, their 
	quartet structure is quite similar as pp-nn quartets exclusively contribute 
	to the electric transition probability. Furthermore, through the analysis of 
	the wave-functions corresponding to the two particle phonons and four 
	particle phonons we noticed that the collective character of these states 
	in not generally pronounced. This conclusion is correlated by low B(E2) 
	transition probabilities calculated values. The relative good agreement 
	of effective charges with respect to the general systematics enables us a reliable prediction
	of B(E2) values.
	
	\section*{Acknowledgments}
	This work was supported by a grant of the Ministry of Research, Innovation and
	Digitization, CNCS - UEFISCDI, project number PN-IV-P1-PCE-2023-0273, within PNCDI IV.
	
	\bibliographystyle{iopart-num}
	\bibliography{references.bib}	
	
\end{document}